# The generalized hardness-intensity diagram for black hole and neutron star X-ray binaries

Chandra B. Singh[1,2,4], David Garofalo[3,4], Kathryn Kennedy[3]

[1] The Raymond and Beverly Sackler School of Physics and Astronomy, Tel Aviv University, Tel Aviv 69978, Israel
[2] South-Western Institute for Astronomy Research, Yunnan University, University Town, Chenggong, Kunming 650500, P.R. China
[3] Department of Physics, Kennesaw State University, Marietta GA 30060, USA
[4] Equal first authors

## Abstract

Over the past half century, X-ray and radio observations of accreting neutron stars and stellar mass black holes have yielded a rich observational picture with common features including state transitions and jet formation, but also sharp differences. While black hole X-ray binaries overwhelmingly suppress jets in so-called soft states, accreting neutron stars are less restrictive, with a soft state wind observed in some sources to co-exist with a jet. We propose an explanation for these differences that leads to a generalization of a foundational element, the hardness-intensity diagram of Fender et al (2004). The inverse relation between jets and winds fits into a picture that connects to prograde accretion while the possibility of counterrotation between accretion disk and compact object accounts for observed differences in accreting neutron stars. This picture comes with a surprising twist, which is that neutron stars embody the small-scale analog of FRII quasars, an idea that allows us to complete the scale invariant picture for the jet-disk connection.

## 1. Introduction

Winds and jets are ubiquitous, present in objects that span a wide range of sizes from young stellar objects, X-ray binaries (XRBs), up to active galactic nuclei (AGN). The origin of such outflows remains an active area of research. In the case of XRBs, correlations between radio (mainly associated with ejection) and X-rays (mainly associated with accretion) have been observed from black holes (Hannikainen, Hunstead, & Campbell-Wilson 1998; Brocksopp et al. 1999; Remillard & McClintock 2006 & references therein; Fender, Homan & Belloni 2009 & references therein) and neutron stars (e.g. Migliari & Fender 2006). Based on accretion properties, black hole X-ray binaries (BHXRBs) are further classified as transient and persistent sources (McClintock

& Remillard 2006) and neutron star X-ray binaries (NSXRBs) with relatively low magnetic fields are classified as atolls and Z-sources (e.g. van der Klis 2005).

The hardness-intensity diagram (Fender, Belloni & Gallo 2004) forms the foundation of our understanding of the disk-jet connection in BHXRBs. A steady compact jet with Lorentz factor up to a few is associated with the so-called hard X-ray state (e.g., Tetarenko et al. 2019 and references therein) but when the source undergoes transition from this hard state to a softer X-ray state, more powerful, optically thin transient jets with Lorentz factor greater than a few are observed (Mirabel & Rodriguez 1994; Hannikainen et al. 2001). In the so-called soft state, the jet turns off or is much weaker, a phenomenon referred to as jet suppression. The compact jet reappears during the soft to hard transition and the cycle is completed. There are some similarities between BHXRBs and NSXRBs which are highlighted in Migliari & Fender (2006). Both NSXRBs and BHXRBs have steady jets in hard states while transient jets are seen at highest luminosities. Surprisingly, jet suppression in soft states is not observed in all NSXRBs (Migliari et al. 2004; however see Miller-Jones et al. 2010) unlike in BHXRBs (e.g. Tananbaum et al. 1972; Ponti et al. 2012). In NSXRBs, transient jets during hard to soft state transitions tend to be weak and not optically thin (Miller-Jones et al. 2010). Attempts at explaining these observations have produced mixed results. Theory and simulations have long suggested that jets from black holes are produced by tapping the spin energy of the black hole in a magnetohydrodynamic process, via the well-known Blandford-Znajek process (Blandford & Znajek 1977). Fender, Gallo & Russell (2010) and Russell, Gallo & Fender (2013) found no correlation between black hole spin and jet power for BHXRBs in hard states while Narayan & McClintock 2012 and King et al. 2011 found dependence of jet power on the black hole spin for BHXRBs in hard to soft transitions. And similar dependence has been seen for some NSXRBs (Migliari, Miller-Jones & Russell 2011; King et al. 2011). The nature of state transitions in different objects is neither unique nor simple, and much progress has been made in modeling the details (e.g. Koljonen 2018).

In this paper, we fit these observations into an explanation that requires a generalization of the hardness-intensity diagram such that it applies to both accreting neutron stars and stellar mass black holes. We suggest that disk orientation is the crucial element in making sense of the observations and that the hardness-intensity diagram of Fender et al (2004) constitutes the behavior of accreting compact objects that are surrounded by prograde accretion disks. When the retrograde window is taken into account, we have a different phenomenology that allows weaker transient jets and absence of jet suppression. In sections 2 and 3, we show that black holes are likely to be more restrictive in the orientation they allow for their disks, with retrograde accretion limited to a subset of the accreting neutron star population. We present model phenomenology in connection with disc-jet coupling and two new hardness-intensity diagrams in section 4. The first amounts to adding the direction of compact object spin in the hardness-intensity diagram of Fender et al (2004) while the second involves counterrotation between the compact object and disk, which may apply to a small subset of NSXRBs. In sections 5 we apply the phenomenology to classify a number of NSXRBs and BHXRBs whose observational properties make them a poster child for the model, and Section 6 concludes.

## 2. Frame dragging

In this Section our goal is to explore the strength of frame dragging near the location of the Lagrange point (the radial location in the equatorial plane of the accretion disk where the compact object gravity equals that of the donor star) to see if there is a meaningful difference between accreting neutron stars and black holes. What we will conclude is that while frame dragging is negligible where the gas from the companion begins to stream into the compact object Roche lobe, black holes may provide sufficiently strong torques that remove most of the incoming gas angular momentum within the Roche lobe prior to the formation of the disk. This appears possible in black hole Roche lobes for two reasons: 1) black hole Roche lobes are physically larger than neutron star Roche lobes which allows torques to act over a larger region and, 2) the gravitational torque from black holes is stronger than neutron star torques. We will provide estimates showing that the incoming gas has a reasonable chance of losing a substantial fraction of its angular momentum which means it will be affected by frame dragging prior to disk formation. The upshot of this is to motivate the idea that unlike for neutron stars, black hole X-ray binaries form accretion disks that tend to be constrained to rotate in the same direction as the black hole rotation. We acknowledge the speculative nature of this constraint for black holes. Nonetheless, the possibility that retrograde accretion is possible only for accreting neutron stars allows the observations to fit into a simple, coherent, picture.

For a binary with compact object mass $M_c$ and donor star mass $m_d$, such that the distance between the binary star centers is unity, the distance of the Lagrange point from the center of the donor star is obtained as follows (Eggleton 1983).

$$r_L/r_d = 0.49(m_d/M_c)^{2/3} / [0.6(m_d/M_c)^{2/3} + \ln(1+(m_d/M_c)^{1/3})] \ . \qquad (1)$$

In Figure 1 we plot 1- $r_L/r_d$, showing the distance from the compact object of the Lagrange point.

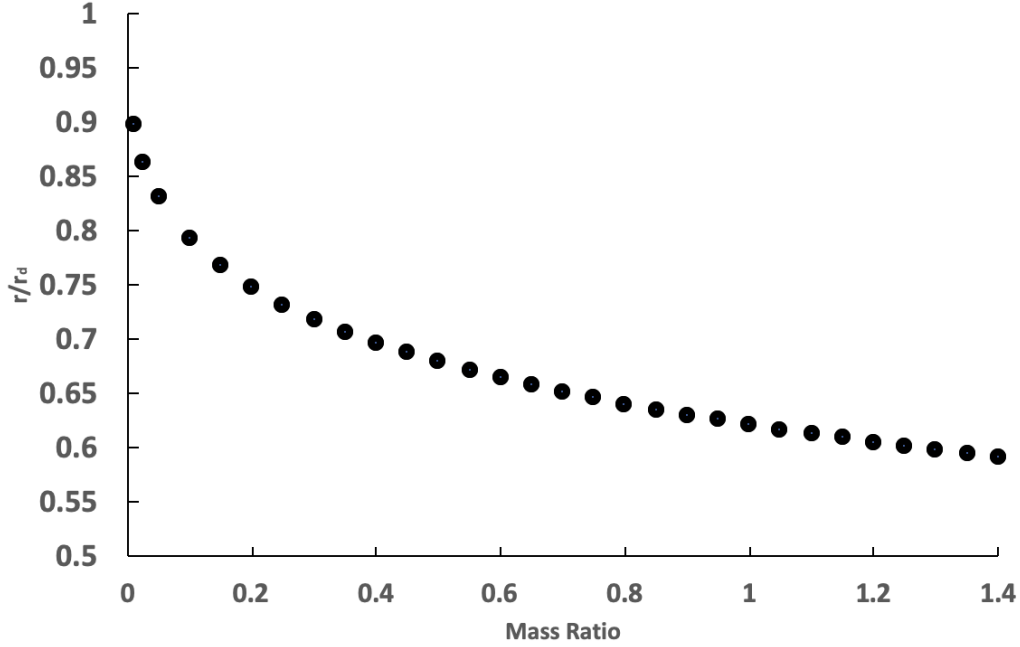

Figure 1: Distance *r* from the compact object to the Roche Lobe radius as a function of mass ratio($m_d/M_c$). $r_d$ is the distance from compact object to donor star. For a given donor star mass, the greater the mass of the compact object, the further away from the compact object is the Roche lobe radius.

Our first goal is to show that frame dragging is greater for black holes than neutron stars despite the fact that both are negligible compared to the angular velocity of inflowing gas at the Lagrange point. From the Hartle-Thorne metric appropriate for neutron stars and the Kerr metric appropriate for black holes, the frame dragging effect at zeroth order in geometrized units is equivalent, and given as follows.

$$\omega = 2M_c a/r^3 \qquad (2)$$

where *a* is the dimensionless spin of the compact object, and *r* is the radial distance from the compact object (Abramowicz et al 2003; Kim et al 2005; Berti et al 2005). If the donor star is about 1 solar mass, $m_d/M_c \sim$ ½ for accreting neutron stars, whereas $m_d/M_c <$ ¼ for accreting stellar mass black holes in principle given the minimum mass for a black hole. In practice it appears a bit lower at 1/8 (Ozel et al 2010). Given the mass ratio for accreting neutron stars, Figure 1 indicates that in the equatorial plane of accretion, the location of the Lagrange point is at about 68% of the distance between the neutron star and the donor star from the neutron star. Given lower values for the mass ratio in accreting black holes, Figure 1 indicates that in the equatorial plane of accretion, the location of the Lagrange point is greater than or equal to about 73% of the distance between the black hole and the donor star from the black hole. As expected, the location of the Lagrange point for the neutron star is closer to the neutron star than for a black hole. Despite the appearance of being deeper in the gravitational potential of the compact object, the frame dragging is actually weaker than in the black hole case as we now show. For this purpose we evaluate a ratio of $\omega_N$ and $\omega_{BH}$ where the subscripts indicate neutron star and black hole frame dragging, respectively. For the minimum black hole mass of just under 4 solar masses, we get the weakest frame dragging effect which is 1.6 times larger than the frame dragging at the location of the Lagrange point for the neutron star case. Hence, the frame dragging ratio satisfies

$$\omega_{BH}/\omega_N > 1.6. \qquad (3)$$

Despite larger black hole frame dragging, the donor star provides gas with non-negligible angular momentum through the Lagrange point, which dominates over frame dragging. We now analyze the possible fate of the gas that enters the black hole Roche lobe using Figure 2 where gas streaming through the inner Lagrange point is shown. As this gas moves to smaller radial distances from the compact object, the gravitational force toward the compact object on the gas increases and the torque about the donor star increases. This torque is negative in the sense that it will decrease the angular momentum of the gas about the donor star. From Figure 2 it is possible to visualize how the magnitude of the decrease in the angular momentum depends on the detailed path of the gas but at a qualitative level we note that if the incoming gas has larger angular momentum about the donor star, it will tend to flow more azimuthally as opposed to radially toward the compact object, and this produces the conditions for a larger total torque and thus a larger total decrease in the angular momentum. If the gas has less angular momentum about the donor star as it streams past $L_1$, it will tend to move more radially than azimuthally toward the compact object and this produces conditions that lead to less total torque and thus less total decrease in angular momentum. Because the compact object-donor star system is isolated, the total angular momentum is conserved which means the decrease in the angular momentum about the donor star for the inflowing gas is compensated by an increase in the angular momentum about the donor star for the compact object. While this constitutes a small effect for the compact object, our goal is to show that it may have a significant effect on the angular momentum retained by the gas and ultimately on its angular velocity about the compact object. The physics of the donor star-compact object interaction can be understood in basic terms as a more complicated version of two rotating disks coming into contact whose axes of rotation are parallel or anti-parallel. As emphasized above, black holes are likely more efficient in their ability to torque the incoming gas and extract its angular momentum because the gravitational force that provides the torque is both larger and can act over a greater range.

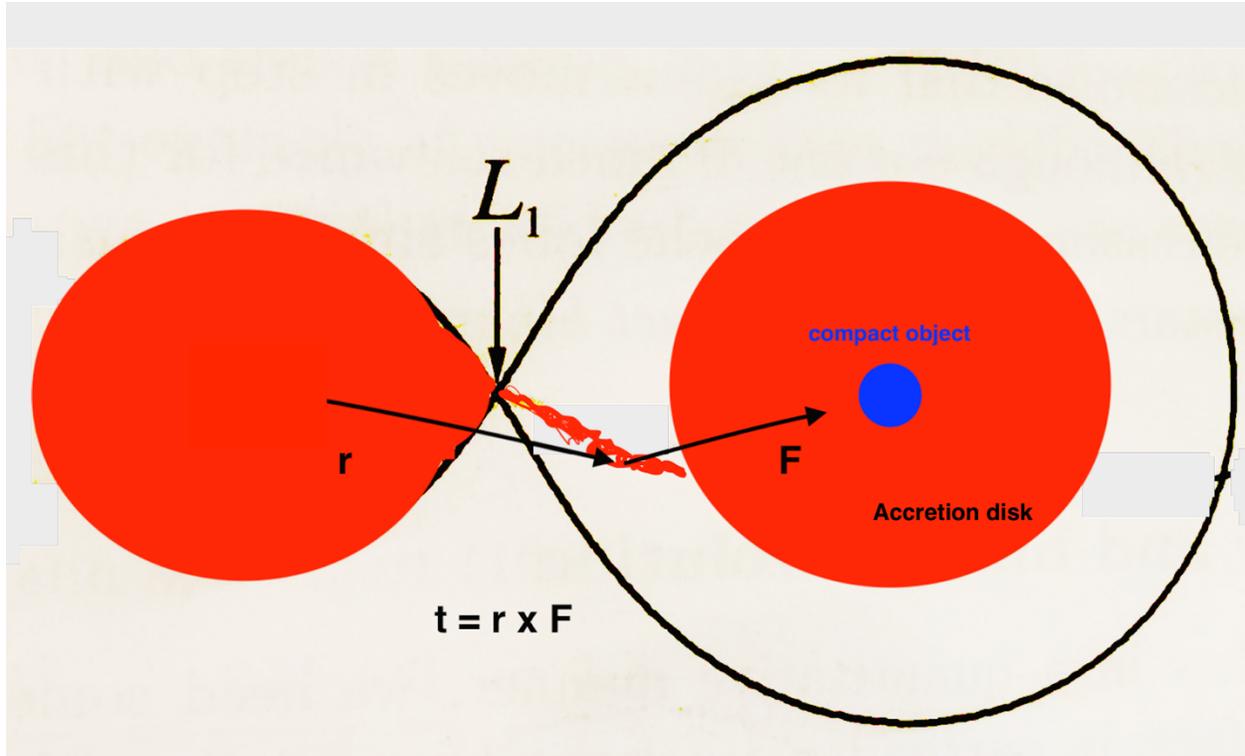

Figure 2: View of Roche lobe overflow accretion onto a neutron star or black hole. The gas that makes up the accretion disk has angular momentum with respect to the donor star which causes it to flow clockwise about the donor star through $L_1$ in this case. But once inside the Roche lobe of the compact object, this gas is eventually dominated by a non-negligible gravitational force toward the compact object indicated as F which produces a negative torque with respect to the donor star. Binary rotation axis is out of the page.

Here we estimate the torque and its ability to extract the angular momentum of the gas at $L_1$ by estimating $F$, the force on a parcel of gas with mass $m$ streaming into the Roche lobe of the compact object. If the gas parcel is along the line connecting the compact object and donor star, the force amounts to a difference in the gravitational forces toward the compact object and toward the donor star as

$$F = m[GM_c/(L-r)^2 - GM_d/r^2] = 4Gm(M_c - M_d)/L^2.$$

Where $L$ is the distance between compact object and donor star and $r$ is the distance of the gas from the donor star. In order to carry out a back-of-the-envelope calculation, we estimate the location of $r$ where the torque is likely strongest to be $r = L/2$. At this value of $r$ the gas is between the Lagrange point and the outer edge of the accretion disk. We also for simplicity assume the angle between $r$ and $F$ is 90 degrees. Under these approximations, the gas parcel does not sit along the line connecting the stars, and the torque about the center of mass of the donor star does not have a contribution due to the gravitational pull of the donor star and is

$$t = r \times F = rF = 4rGmM_c/L^2.$$

The angular momentum of the gas at the Lagrange point is

$$L = m\omega r_1^2 = m\omega L^2/16$$

which comes from estimating the location of the Lagrange point to be $r_1 = L/4$. We again simplify by assuming the gas at the Lagrange point follows a Keplerian angular velocity which is an overestimate, and obtain

$$\omega = [GM_d/r^3]^{1/2} = [64GM_d/L^3]^{1/2}.$$

If we were to choose to evaluate the torque about some other point, the torque would depend on the difference between the stellar masses, which would allow us to notice that only in situations where q < 1 by a factor of at least a few, that a reasonable expectation exists for a significant effect on the angular momentum of the gas at $L_1$. Nonetheless, smaller q allows for a larger total torque on the gas because the Roche lobe of the compact object is larger and therefore a larger distance between $L_1$ and the radial location of the outer edge of the accretion disk for the torque to grow large enough to affect the angular momentum of the incoming gas. In other words, for binaries where q is order unity, there is little opportunity for the angular momentum of the inflowing gas to be removed. The reference point about which we carry out our analysis does not matter. If we choose a random point, the greater torque for the black hole case emerges from the difference in the masses between compact object and donor star (smaller q). If, instead, we choose to evaluate the torque about the donor star, the torque will only depend on compact object mass which is of course larger for black holes. We choose to carry out the analysis following the latter approach.

We can now proceed to an estimate of the timescale necessary for the torque to remove the angular momentum of the gas streaming through $L_1$. This amounts to an integral of the torque over the time it acts but we further simplify by assuming the torque is constant. And to evaluate whether that time is reasonable we will estimate and compare to the free fall time for the gas to travel from the Lagrange point to the accretion disk. In other words, we connect the constant torque $\tau$, the time it acts on the gas T, and the angular momentum at $L_1$, via

$$\tau T = L$$

which gives

$$T4(L/4)GmM_c/L^2 = m\omega r^2 = m[64GM_d/L^3]^{1/2} L^2/16 = m[GM_d L]^{1/2}/2$$

which in turn simplifies to

$$TGM_c/L = [GM_d L]^{1/2}/2.$$

Solving for the time we get

$$T = [GM_d L^3]^{1/2}/[2GM_c]$$

which further simplifies to

$$T = [M_d L^3/G]^{1/2}/[2M_c]$$

and if we assume $L$ is about the Sun-Mercury distance, the donor star is 1 solar mass, while the compact object is about 3 solar masses, we get T=2.4 days which compared to the free-fall timescale ($T_{ff} = (r^3/GM)^{1/2}$) of 8.5 days for a Mercury-Sun distance, suggests there is ample time to extract the angular momentum at $L_1$. If we consider larger compact object masses, we get free-fall timescales that are smaller but for the range of black hole masses in X-ray binaries this timescale is reduced by a small factor and remains on the order of days. For example, if the black hole mass is 10 solar masses, the required time T is reduced to 0.73 days while the free-fall time is about 4.5 days.

The results of our back-of-the-envelope calculations suggest that black holes may be capable of extracting the angular momentum of the incoming gas. Of course, whether this occurs or not depends on the detailed path that inflowing gas follows as it streams into the black hole Roche lobe. This is therefore quite speculative. However, if most of the angular momentum is in fact extracted, frame dragging must have an impact. As Figure 1 shows, neutron stars accrete gas from Lagrange points that are closer to the neutron star and with weaker torques of the kind shown in Figure 2. As a result, they are less likely to strongly impact the angular momentum of the incoming gas. As a result, the gas flowing through the Lagrange point into the neutron star Roche lobe will flow into an accretion disk in the prograde or retrograde direction compared to the neutron star rotation as determined by the angular momentum of the gas at $L_1$. In other words, neutron star frame dragging is always neglible even close to the neutron star. Hence, neutron stars are not constrained to corotating disks as their black hole counterparts. The conditions that allow the retrograde disk to stably form require $J_d < 2J_c$ where $J_c$ is the angular momentum of the compact object or neutron star and $J_d$ that of the disk (King et al 2005). This is even less restrictive than in the supermassive black hole case where the disk is fed by an entirely different process and whose size is less constrained.

In summary, for accreting stellar mass black holes, we have provided an estimate for the torque produced by the black hole on the incoming gas arguing that it may be effective enough to extract its angular momentum, thereby allowing frame dragging deep inside the black hole Roche lobe to force the gas to flow into co-rotation with the black hole. Neutron stars, on the other hand, produce weaker frame dragging as well as smaller Roche lobes so their torques on the gas will extract less angular momentum from it and therefore are unlikely to force the donor star gas into co-rotation. These ideas motivate the possibility that retrograde and prograde accretion should occur in neutron stars, unlike in black holes.

## 3. Accreting neutron stars as small-scale FRII quasars

In the previous section we showed that torques may develop in black hole Roche lobes that remove much of the angular momentum of the incoming gas such that frame dragging close to the black hole may come into play in determining the final fate of the gas in the disk. Because this seems to be less likely for neutron stars, their disks will develop co-rotation or counterrotation as determined by the angular momentum of the gas at $L_1$. We note that evidence of spin down or retrograde accretion in Be/X-ray binaries has been explored in wind fed systems around neutron stars (Christodoulou, Laycock & Kazanas 2018). However, we do not explore these systems because their magnetic fields tend to be much stronger than those that we envision to be compatible with our jet-disk connection and hard-soft transition picture.

Having motivated the idea that counterrotation may occur in low mass X-ray binary neutron stars, allows us to apply the ideas concerning retrograde accretion around supermassive black holes to neutron stars. The bottom line is that retrograde accretion allows for both powerful jets as well as disk winds (Garofalo, Evans & Sambruna 2010). The reason for this has to do with larger gap regions between the compact object and the inner edge of the accretion disk at the innermost stable circular orbit (ISCO) for retrograde accretion. The ISCO for retrograde accreting neutron stars comes from a study of the stability of circular orbits in the Hartle-Thorne metric and depends on the mass M, angular momentum j, and multipole moment q as (Abramowicz et al 2003)

$$R_\pm = 6M [1 -+ j(2/3)^{1.5} + j^2(97.09 - 240 \ln 1.5) + q (-97.13 + 240 \ln 1.5)]$$

where the + sign refers to co-rotation and the – sign to counterrotation.

The quantitative details for the ISCO depend on the equation of state and have been addressed in Berti et al (2005). We show results for the ISCO in Figure 3. Because the gap region is at most about twice the radius of the neutron star, it is not as large as it can be in the black hole case where the ISCO may extend as far out as 9 gravitational radii for highest black hole spin. In order to further understand jet formation and jet suppression in black holes versus neutron stars, we should emphasize that the function of the gap region in this respect is threefold: 1) It allows for a greater accumulation of magnetic flux on the compact object compared to systems with smaller gap regions. 2) The large flux bundle threading the compact object allows for a greater bend of the magnetic field lines threading the disk which leads to an enhancement in the Blandford-Payne jet. 3) The absence of near compact object accretion material (i.e. a gap region) limits the amount of energy that is reprocessed further out in the disk. Since a greater amount of reprocessed energy leads to disk winds that suppress jets, an absence of such energy makes jet suppression more difficult. The details of this jet-disk connection can be found in Garofalo, Evans & Sambruna (2010). We note that at least qualitatively the weaker gravitational potential in which neutron star ISCOs find themselves implies that smaller gap regions may still be compatible with absence of jet suppression. In other words, neutron stars do not need as large a gap region to avoid jet suppression compared to black holes. What needs highlighting is not the existence of jets in accreting neutron stars, but the presence of jets from neutron stars that are in soft states. If we were to flip the rotation of the neutron star so that the disk becomes prograde accreting, the ISCO would either disappear (left panel of Figure 3) or would move inward toward the stellar surface to under 14 km (right panel of Figure 3). This decrease in the gap region would be associated with a decrease in the magnetic flux accumulated onto the neutron star, a weaker disk jet, and stronger jet suppression, thereby reducing jet power.

The nature of jet suppression, therefore, is contingent on the size of the gap region, which progressively decreases as the spin increases in the prograde direction. Because intermediate prograde thin disks have larger gap regions compared to high spinning prograde systems, they suffer less jet suppression. This has been explored in the supermassive black hole case for the so-called FRI radio quasars (Kim et al. 2016) and allows one to appreciate where weak jets in soft states might emerge in the model for black hole X-ray binaries, i.e. this occurs for systems that do not possess the highest black hole spins. Our scale invariant explanation for jets in soft states for neutron star low mass X-ray binaries allows us to make close contact with the black hole case. Note also the possibility of observational evidence for retrograde configurations in accreting stellar mass black holes that would invalidate our ideas (Morningstar, Miller, Reis, Ebisawa 2014; Reis et al 2013; but see Garofalo, Kim & Christian 2014).

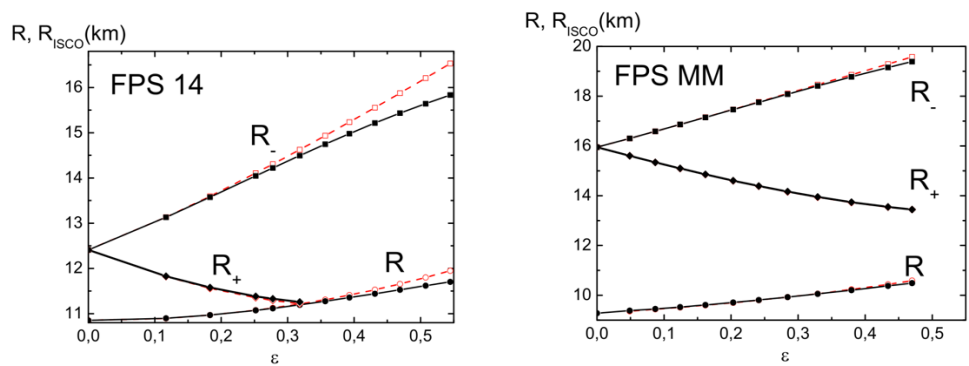

Figure 3: Neutron star radius (labeled R and indicated with circles) and ISCO values as a function of rotation parameter ε for different equations of state FPS14 and FPS MM from Berti et al (2005). Corotating ISCO is labeled $R_+$ (diamonds) while counterrotating ISCO

is labeled R₋ (squares). Note that the corotating ISCO does not always exist. Red refers to the values from the Hartle-Thorne metric.

## 4. Generalized hardness-intensity diagrams

In this section we apply the high-spin prograde/thin disk/jet suppression and high-spin retrograde/thin disk/no jet suppression idea to produce a generalization of the hardness-intensity diagram. To this basic idea we add additional physics described briefly. Because strong radiative disk winds suppress jets in this model, retrograde configurations are the weakest jet suppressers while prograde configurations are the strongest jet suppressers. But strong radiative winds are a feature of thin disks. In hard states, therefore, jet suppression fails due to disk thickness (Garofalo, Evans & Sambruna 2010; Garofalo & Singh 2016). And when a thick disk collapses to a thin disk in prograde configuration, the radiative wind from the incoming thin disk serves to momentarily collimate the material that is already trapped onto the inner field lines, producing a transitory burst jet, followed by jet suppression once material no longer makes it onto the inner field lines. In the retrograde configuration, on the other hand, the collapse from the hard state of the thick disk into a thin disk produces weaker radiative disk wind, which then fails to have the same effect as the prograde counterpart and no strong transitory jet is produced. The system simply transitions to the soft state without strongly affecting its jet, which persists (Garofalo, Evans & Sambruna 2010; Garofalo & Singh 2016). Interestingly, the coupling of the Blandford-Znajek and Blandford-Payne jets in the context of the Reynolds Conjecture, also produces a weaker jet dependence on spin in the prograde configuration compared to the Blandford-Znajek mechanism in isolation (Garofalo, Kim & Christian 2014).

We apply the phenomenology described above to BHXRBs and NSXRBs and show that we are led to a generalization of the hardness-intensity diagram of Fender, Belloni & Gallo (2004). Our generalization comes in the form of two Figures, 4 and 5. Figure 4 constitutes a small addition to the diagram of Figure 7 of Fender, Belloni & Gallo (2004) by adding the rotation of the black hole relative to that of the accretion disk. Our claim as motivated by the analysis above is that the incoming material, whether it is from a wind or from Roche lobe overflow, will be forced into co-rotation with the black hole. Hence, the hardness-intensity diagram of Figure 4 applies to all stellar mass black holes that experience state transitions as well as to counterpart prograde accreting NSXRBs. Two features appear on this diagram that need emphasis as they are absent on the hardness-intensity diagram for counterrotating compact objects. First, the existence of a jet line, demarcating the region where jets exist and where they are absent. Second, the jet Lorentz factor receives a boost in the state transition. Both of these features emerge in the model as a result of the transition from a thick disk to a thin one in a way that hinges on the closeness of the disk inner edge to the compact object. In Figure 5 we produce a hardness-intensity diagram for counterrotating systems (which applies only to a subset of the NSXRBs) whose inner thin disk regions live further away from the compact object. As a result of this feature, counterrotating systems that experience state transitions are much less capable of suppressing jets and generating strong transitory burst jets. Note the presence of a jet in the soft state (lower left) unlike in Figure 4.

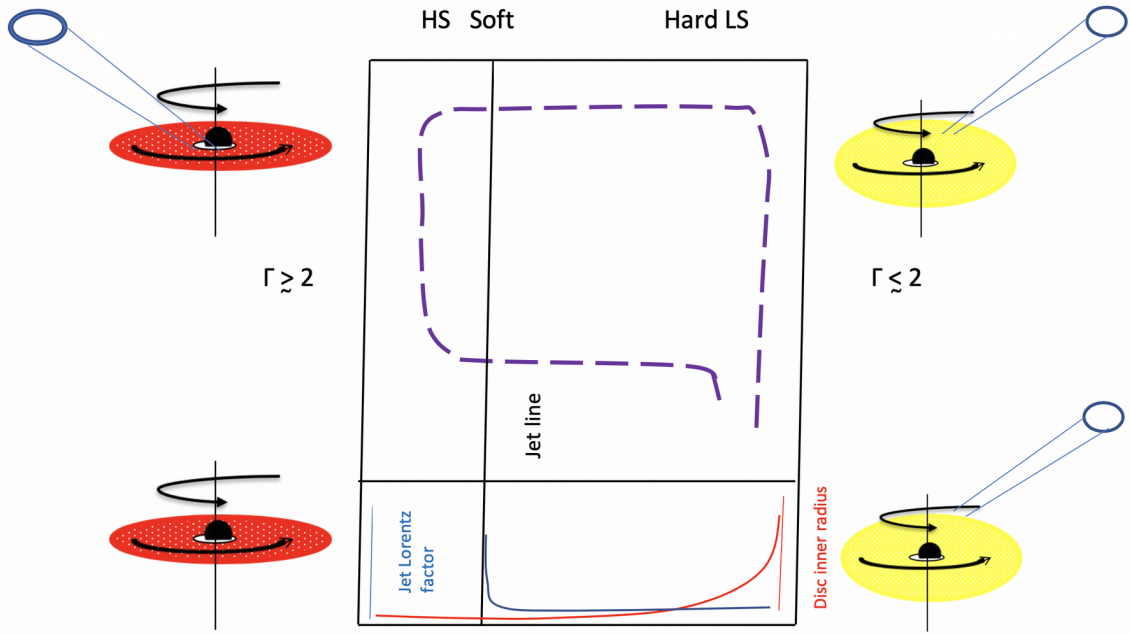

Figure 4: Hardness-intensity diagram for X-ray binaries in prograde accretion configurations (includes all BHXRBs and some NSXRBs). The diagram is designed to be equivalent to Figure 7 in Fender, Belloni & Gallo (2004) except for the additional feature of the rotation of the compact object which shares the angular momentum direction with its disk. Yellow represents hard X-ray states while red soft ones. Unlike in Fender, Belloni & Gallo (2004), for simplicity we have not added additional features to our jets.

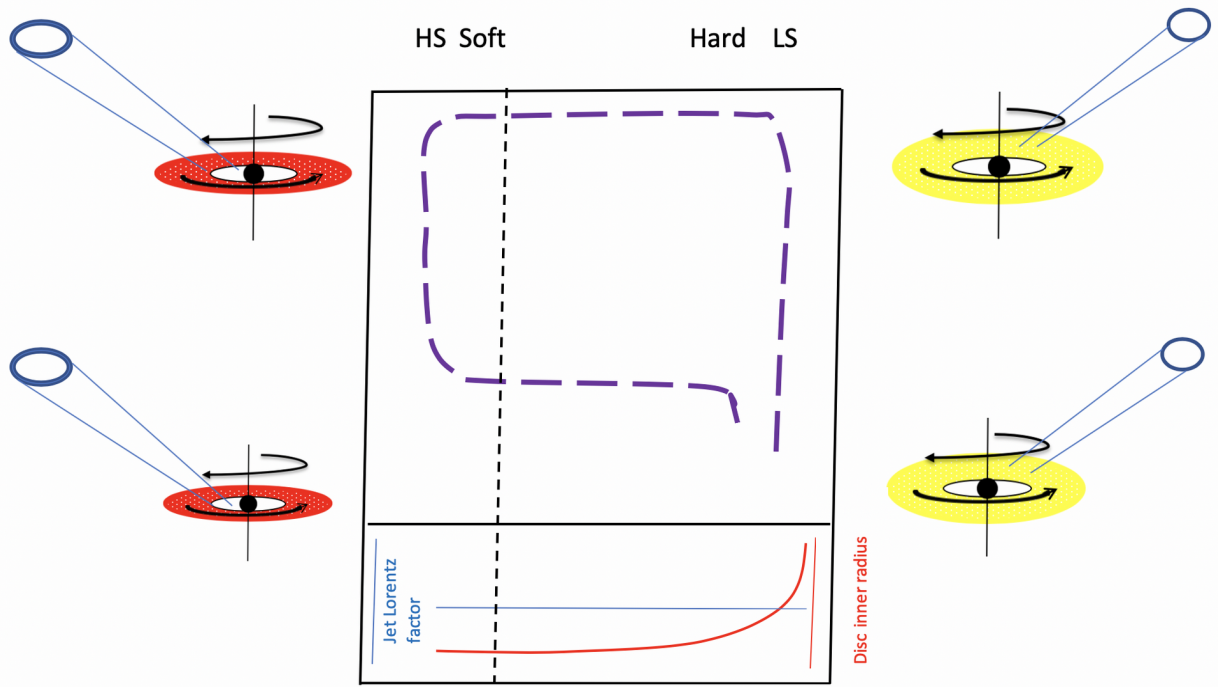

Figure 5: Hardness-Intensity diagram for X-ray binaries in retrograde accretion configuration (includes only a subset of NSXRBs). Because of the large gap region, no jet suppression occurs in the soft state. While the Lorentz factor is not prescribed since neutron stars are different species of accreting compact objects, the model does not identify a mechanism that can affect the jet in the state transition. Because the disk inner radius is larger for retrograde accretion, the red line is shifted further up compared to its Figure 4 counterpart. Although no jet line exists for such objects, the dashed line shows where it is in Figure 4.

## 5. Individual NSXRBs and BHXRBs

In this section our goal is to highlight observed objects that in the model fit nicely in the retrograde/prograde, soft/hard dichotomies. Atoll NSXRBs have accretion rates in the range 0.01-0.5 of the Eddington accretion rate $(dM/dt)_{Edd}$ while Z-sources are found to have accretion rates above 0.5 $(dM/dt)_{Edd}$. The Atoll source Aql X-1 and the Z-source Sco X-1 have their jet radio emission quenched during the soft states. This quenching or suppression of the jet is the hallmark of prograde accretion (Figure 4). On the other extreme, two atoll sources 4U 1820-30 and Ser X-1 produce jet radio emission during the soft state (Migliari et al. 2004) which is the hallmark of retrograde accretion (Figure 5) and a feature that is not shared by any BHXRBs. The ability to suppress jets in soft states is most pronounced in the model at high prograde spin, while a soft state in a high spin retrograde system experiences weakest jet suppression. In fact, the model predicts a correlation between increasing spin (i.e. in the direction of high retrograde to high prograde) and jet suppression. FRII quasars are the AGN analog of the lower left system of Figure 5 but with a black hole replacing the neutron star and a merger replacing the donor star as the feeding mechanism (Garofalo, Evans & Sambruna 2010; Garofalo & singh 2016). Radiatively inefficient accretion states (i.e. hard states), on the other hand, have no ability to suppress jets which is why both orientations produce jets in such accretion states. These are the extremes that make sense of Figures 4 and 5.

But as we consider compact objects whose spins are not high and accretion rates that are intermediate, differences will appear that influence the presence, strength, and type (optically thin/optically thick) of jet and disk wind which competes with it. For example, less bright thin disks could result from lower accretion rates but also from counterrotation which has lower disk efficiency. Although not dominant in NSXRBs, some optically thin jets are seen. For BHXRBs, optically thin jets appear in hard to soft state transitions which in the model emerges from the collapse to a thin disk and the increasing radiative efficiency of the disk as the inner edge moves closer toward the compact object. These are physical manifestations of hard to soft transition in a high spinning prograde accreting compact object. The presence of such jets in some NSXRBs suggests therefore, high prograde spin does occur in neutron stars, but the paucity in the numbers can be understood in this model in that NSXRBs have a larger parameter space to choose from compared to BHXRBs.

Our model raises some issues. Retrograde accretion is short-lived in Eddington-limited accreting black holes and is perhaps a bit longer in neutron stars due to smaller ISCO values, namely less than 10 million years. Yet, magnetic fields in the low mass NSXRBs that we are modeling as retrograde accretors are orders of magnitude lower than those in high mass NSXRBs and pulsars (Manchester et al. 05). The problem is that the lower magnetic fields are thought to result from prolonged accretion periods that lead to field decay on timescales of 10 million years (Urpin & Geppert 1995; Bisnovatyi-Kogan & Komberg 1974; Bhattacharya 2011) which are inconsistent with retrograde accretion. The ideas presented in this paper, therefore, require some mechanism that reduces nascent magnetic fields of $10^{12}$ Gauss to $\sim 10^8$ Gauss that is not accretion based and that in fact can operate prior to Roche lobe overflow accretion onto the neutron star. Ambipolar diffusion has been recently investigated (Cruces, Reisenegger & Tauris, 2019; see also Tauris & Konar 2001) as such a mechanism. Despite the possibility of a decay in the magnetic field prior to accretion, neutron star spin emerges with low values. Hence, we seem to be constrained to a model in which retrograde accretion in low mass NSXRBs threaded by $\sim 10^8$ G magnetic fields is possible but such that the spin value is not high. This means that unlike in the black hole case, retrograde accreting neutron stars have a greater ability to maintain the jet in a soft state compared to black holes. A scale-invariant estimate for the Eddington-limited accretion timescale for spinning down a rapidly spinning compact object is just under $10^7$ years and to spin it up to high spin again would take about $10^8$ years. Given that neutron stars seem constrained to produce low retrograde spin, the rela-tive timescales compared to the prograde regime is even smaller. In short, while retrograde accretion constitutes a small subset of active galaxies in our paradigm, it seems to amount to an even smaller fraction of the total accreting subgroup in stellar mass compact objects.

Transient BHXRBs that more cleanly fit into the model include GX 339-4 with $a<0.9$ (Kolehmainen & Done 2010; Reis et al. 2008; Garcia et al. 2015; Parker et al. 2016), GRO J1655-40 with $a=0.7\pm0.1$ (Shafee et al. 2006; Reis et al. 2009), H 1743-322 with $a = 0.2\pm0.3$ (Steiner, McClintock & Reid 2012) and persistent sources like GRS 1915+105 with $a> 0.98$ (McClintock et al. 2006; Miller et al. 2009). These objects produce steady hard state jets that are suppressed in soft states and therefore fit as high spinning prograde systems (Figure 4).

Whether transient or persistent, atoll or Z-source, NSXRBs and BHXRBs display a variety of observational features that are difficult to incorporate into a single coherent framework. We have shown how to fit them into more physically fundamental categories. While we have picked objects among the two species of accreting compact objects that fit most nicely into the paradigm, we have also argued that a natural space exists to make sense of observations of mixed properties. The hope is that observers will take up the challenge.

## 6. Conclusions

Despite the speculative nature of our constraint, we have motivated the idea that disk orientation may be constrained only for accreting black holes and the realization that retrograde accretion may explain the behavior of some accreting neutron stars has led us to a generalized hardness-intensity diagram according to which disk orientation makes a difference to state transitions in terms of the presence or absence of jet suppression and transitory burst jets. We have suggested that the hardness-intensity diagram should be divided into prograde and retrograde versions as the fundamental units by which to understand accreting stellar mass compact objects. By opening up the retrograde window we also completed a scale-invariant framework for the jet-disk connection suggesting that some of the most powerful AGN families share properties not with small-scale black holes but with accreting neutron stars. A natural extension of the ideas presented in this work involves accreting white dwarfs whose soft states are coupled to substantial radio emission.

**Acknowledgments**

We would like to thank anonymous referee whose suggestions helped us to improve the quality of manuscript.